\documentclass[aps,prl,showpacs,groupedaddress]{revtex4}
\usepackage[dvips]{graphicx}
\usepackage{dcolumn}

\begin{document}

\title
{Conductivity in glass phases of disordered granular superconductors in magnetic fields
}

\author{Ryusuke Ikeda}

\affiliation{%
Department of Physics, Kyoto University, Kyoto 606-8502
}

\date{\today}

\begin{abstract}
The electric conductivities in glass phases of three-dimensional (3D) granular superconductors in magnetic fields are examined based on a disordered quantum Josephson-junction array. Contrary to a recent argument of a glass phase with {\it metallic} response, a correct inclusion of an Ohmic dissipative dynamics always leads to a glass phase with divergent dc conductivity. With no dissipative term, a metallic glass phase is obtained irrespective of the range of correlation of quenched disorder, i.e., even in the so-called Bose-glass phase with experimentally vanishing resistivity. 
\end{abstract}


\maketitle

\section{I. Introduction}

Recently, it was argued based on a calculation of the conductivity of two-dimensional (2D) disordered Josephson-junction arrays \cite{DP,PD} that a glass phase peculiar to disordered granular superconductors, called as a phase glass (PG), is not a superconducting but metallic phase with a nonvanishing resistance at low enough $T$. It is quite important to judge this surprising argument correctly in order to clarify whether the resistivity data suggestive of the presence of an intermediate quantum "bose metal" phase \cite{Aharon,Vicente} are intrinsic or not. Subsequently, the argument \cite{DP,PD} of the glass phase with a nonvanishing resistance was straightforwardly extended to the 3D case and the systems under nonzero magnetic fields, and thus, the so-called vortex-glass (VG) phase \cite{FFH} resulting from a point-like quenched disorder at nonzero temperatures was argued \cite{WP} to be also a metal phase, contrary to the conventional wisdom. \cite{Feigel} However, their argument seems to be based essentially on their assumption of nondissipative bare dynamics of the phase field $\theta$. The neglect of Ohmic dissipation is not acceptable in discussing, at least, the VG phase at nonzero temperatures. 
\begin{figure}[c]
\caption{Schematic {\it mean-field} $H$-$T$ phase diagram of a disordered granular superconductor. The dashed curve implies the microscopic $H_{c2}(T)$ curve for a {\it single} grain.} \label{fig.1}
\end{figure}

In the present paper, we show that inclusion of an Ohmic dissipative dynamic term always leads to the conventional picture that any glass phase of granular superconductors in nonzero fields is a superconducting phase with vanishing dc resistivity in the plane perpendicular to a uniform magnetic field ${\bf H} \parallel {\hat z}$. In addition, we point out that, with no dissipative dynamics of $\theta$, even the so-called Bose glass (BG) phase created by a line-like (columnar) disorder parallel to ${\hat z}$ becomes a metal phase. Based on the experimentally well-accepted fact that the BG phase is superconducting, we argue that the argument \cite{DP,PD} of the metallic (nonsuperconducting) VG phase based on the use \cite{DP,PD,WP} of nondissipative dynamics has no foundation acceptable physically. These results are consistent with those of the analysis \cite{RIsub} from the vortex liquid regime based on the Gaussian glass fluctuation. 
In contrast, the authors of Ref.\cite{DP,PD} have not studied consistently the regime in nonzero temperatures just above a glass transition. 

\section{II. Conductivity in glass phase due to uncorrelated disorder}

We start from the hamiltonian 
\begin{equation}\label{eq:1} 
{\cal H}_\theta = \alpha \sum_{\bf r} \biggl(-{\rm i} \frac{\partial}{\partial {\hat \theta}_{\bf r}} \biggr)^2 - \sum_{{\bf r}, {\bf \mu}} \frac{J_{\bf \mu}({\bf r})}{2} \, {\rm cos}({\hat \theta}_{\bf r} - {\hat \theta}_{{\bf r}+{\bf \mu}a}), 
\end{equation}
describing a Josephson junction array with a charging energy $2 \alpha$ on each grain, where ${\bf r}$ denotes the coordinate of each site (i.e., grain), ${\bf \mu}$ is the unit vector pointing to possible nearest-neighbor directions, $a$ the lattice constant of the simple cubic or square lattice, and ${\hat \theta}_{\bf r}$ is a phase operator on the grain at ${\bf r}$. The model may be extended to a more general one including effects of possible dissipation on each grain and of electromagnetic fields. The most straightforward method of performing this is to express the model (\ref{eq:1}) into the corresponding quantum action 
\begin{eqnarray}\label{eq:2}
{\cal S}\! &=& \! - \! \int^\beta_0 \! d\tau \! \sum_{{\bf r}, {\bf \mu}} \biggl[ \frac{J_{\bf \mu}({\bf r})}{4} \exp[ \, {\rm i}(\theta_{\bf r}(\tau) - \theta_{{\bf r}+{\bf \mu}}(\tau) - e^* A_{{\rm ex}, {\bf \mu}}({\bf r}) - e^* \delta A_{\bf \mu}({\bf r}, \tau)) \, ] 
+ c.c. \biggr] \\ \nonumber 
&+& \int^\beta_0 d\tau \sum_{\bf r} \frac{1}{4 \alpha} \biggl(\frac{\partial \theta_{\bf r}(\tau)}{\partial \tau} \biggr)^2
\end{eqnarray}
in the unit $\hbar=c=1$, where $\beta=1/T$, $e^*$ is the Cooper-pair charge, $A_{{\rm ext},{\bf \mu}}$ denotes the line-integral of an external gauge field over a single bond in the ${\bf \mu}$-direction, and $\delta A_{\bf \mu}(\tau)$ is the corresponding gauge disturbance introduced for obtaining the conductivity in the ${\bf \mu}$-direction. 
Further, 
\begin{eqnarray}\label{eq:3}
{\cal S}_0(\theta) &=& \int^\beta_0 d\tau \sum_{\bf r} \frac{1}{4 \alpha} \biggl(\frac{\partial \theta_{\bf r}(\tau)}{\partial \tau} \biggr)^2
\end{eqnarray}
is the action corresponding to the charging energy, i.e., the first term of eq.(1). 

Note that the dissipative (last) term of eq.(\ref{eq:2}) is expressed as 
\begin{equation}\label{eq:4}
{\cal S}_{\rm dis} = \beta^{-1} \sum_{\bf r} \sum_\omega \frac{\nu}{2} |\omega|  |\Phi_{\bf r}(\omega)|^2,
\end{equation}
where $\Phi_{\bf r}(\omega)$ is the Fourier transform of 
\begin{equation}\label{eq:5}
\Phi_{\bf r}(\tau) = \exp({\rm i} \theta_{\bf r}(\tau)). 
\end{equation}
That is, eq.(\ref{eq:4}) is nothing but the familiar dissipative term, written in the phase-only approximation, in the time-dependent Ginzburg-Landau model. 

Hereafter, let us proceed to rewriting the action ${\cal S}$ into a form convenient for a field-theoretical method. Further, a point-like (uncorrelated) disorder will be assumed in this section. A quenched disorder in the system is incorporated into a randomness of $J_{\bf \mu} = J_{-{\bf \mu}}^*$ with a nonzero real mean $J_0$, i.e., ${\overline {J_{\bf \mu}}}=J_0 > 0$,  and a Gaussian distribution ${\overline {(J_{\bf \mu}({\bf r}_1) - J_0) (J_{-{\bf \mu}}({\bf r}_2) - J_0)}} = 4 \delta_{{\bf r}_1, {\bf r}_2} J^2/d$, where $d$ is the space dimension. These relations may be regarded as being due to a {\it random} gauge field $a_{\bf \mu}$ defined by $J_{\bf \mu}({\bf r}) - J_0 \propto \exp[{\rm i} a_{\bf \mu}({\bf r})]$. 
The free energy $F = - \beta^{-1} {\rm ln} Z$ will be expressed in terms of the replica trick as $F = - \beta^{-1} (Z^n - 1)/n$ in $n \to +0$ limit. The averaged replicated partition function ${\overline {Z^n}}$ is given by 
\begin{equation}\label{eq:6}
{\overline {Z^n}} = {\overline Z^n_0} \, < \exp(- {\cal S}_f - {\cal S}_g) 
>_0, 
\end{equation}
where $Z_0$ is the partition function of ${\cal S}_0$, $< \, \, \, >_0$ denotes the ensemble average on $\sum_{1 \leq a \leq n} {\cal S}_0(\theta^{(a)})$, and 
\begin{eqnarray}\label{eq:7}
{\cal S}_{\rm int} &=& - \sum_{a=1}^n \sum_{{\bf r}, {\bf \mu}} \int_0^\beta \! d\tau \frac{J_0}{2} \, {\rm cos}( e^* \delta A_{\bf \mu}(\tau) + e^* A_{{\rm ex},{\bf \mu}}({\bf r}) - \theta_{\bf r}^{(a)}(\tau) + \theta_{{\bf r}+{\bf \mu}a}^{(a)}(\tau) ) \nonumber \\
&-& \frac{1}{4d} \int d\tau \int d\tau' \sum_{a,b} \sum_{{\bf r}, {\bf \mu}}  J^2 \, \cos \, ( \, e^* [ \, \delta A_{\bf \mu}(\tau) 
- \delta A_{\bf \mu}(\tau') \, ] \nonumber \\
&+& \theta_{\bf r}^{(a)}(\tau) - \theta_{\bf r}^{(b)}(\tau') - \theta_{{\bf r}+{\bf \mu}a}^{(a)}(\tau) + \theta_{{\bf r}+{\bf \mu}a}^{(b)}(\tau') \, ). 
\end{eqnarray}
Before proceeding further, ${\cal S}_f$ will be rewritten in the form \cite{RIsub} 
\begin{eqnarray}\label{eq:8}
{\cal S}_f &=& {\rm const}. - d J_0 \beta^{-1} \sum_\omega \sum_{\bf r} \Phi^*_{\bf r}(\omega) \biggl( 1 - \frac{\nu}{2 d J_0} |\omega| + \frac{1}{2d} \sum_{\bf \mu} {D}_{\bf \mu}({\bf r}) \cdot {D}^*_{\bf \mu}({\bf r}) \biggr) \Phi_{\bf r}(\omega) 
\\ \nonumber 
&\simeq& {\rm const}. - d J_0 \beta^{-1} \sum_\omega \, \biggl( 1 + \frac{\nu |\omega|}{2 d J_0} \biggr)^{-1} \sum_{\bf r} \Phi^*_{\bf r}(\omega) \nonumber \\
&\times& \biggl( 1 + \frac{1}{2d} \sum_{\bf \mu} {D}_{\bf \mu} ({\bf r}) \cdot {D}^*_{\bf \mu}({\bf r}) \biggr) \Phi_{\bf r}(\omega) \\ \nonumber
\end{eqnarray}
for the cubic or square lattice in $d$-dimension. Equation (9) is valid up to the lowest order in $ \nu |\omega|/J_0$ and in the gauge-invariant gradient ${D}_{\bf \mu}$ on the lattice \cite{Kleinert} accompanied by the gauge field ${\bf A}_{\rm ex} + \delta {\bf A}(\tau)$. 
Then, by introducing the conventional SC order parameter $\psi^{(a)}({\bf r}, \tau)$ and the PG order parameter $q^{(ab)}_{\bf r}(\tau_1,\tau_2)=(q^{(ba)}_{\bf r}(\tau_2, \tau_1))^*$, ${\overline Z^n}$ becomes 
\begin{equation}\label{eq:9}
\frac{\overline {Z^n}}{\overline {Z^n_0}} = \int {\cal D}\psi^{(a)} {\cal D}(\psi^{(a)})^* {\cal D} q^{(ab)} \exp(-{\cal S}_{\rm eff}), 
\end{equation}
where 
\begin{eqnarray}\label{eq:10}
{\cal S}_{\rm eff} \! &=& \! \sum_{\bf r} \biggl[ \int \! \! d\tau_1 \! \! \int \! \! d\tau_2 \sum_{a,b} \frac{J^{-2}}{2} q^{(ab)}_{\bf r}(\tau_1, \tau_2) q^{(ba)}_{\bf r}(\tau_2, \tau_1) + \frac{\beta^{-1}}{4d} \sum_{\omega,a} \biggl(1 + \frac{\nu}{2d J_0} |\omega| \biggr) |\psi^{(a)}_\omega ({\bf r})|^2 \\ \nonumber 
&-& {\rm ln} \biggl( \biggl< T_\tau \exp \biggl(\frac{\sqrt{J_0}}{2} \int d\tau \sum_a \Phi_{\bf r}^{(a)}(\tau) \biggl( 1 + \sum_{\bf \mu} \frac{{D}_{\bf \mu} \cdot {D}_{\bf \mu}^*}{2d} \biggr)^{1/2} 
(\psi_{\bf r}^{(a)}(\tau))^* \nonumber \\
&+& \! \! \! \frac{1}{2} \int d\tau_1 d\tau_2 \sum_{a,b} \Phi_{\bf r}^{(a)}(\tau_1) (\Phi_{\bf r}^{(b)}(\tau_2))^* \biggl( 1 + \sum_{\bf \mu} \frac{{\tilde {D}}_{\bf \mu} \cdot {\tilde {D}}_{\bf \mu}^*}{2d} \biggr)^{1/2} q_{\bf r}^{(ba)}(\tau_2, \tau_1) + {\rm c.c.} \biggr) \biggr>_0 \biggr) \biggr],  
\end{eqnarray}
\begin{equation}\label{eq:14}
\psi(\tau) = \beta^{-1} \sum_\omega \psi_\omega e^{-{\rm i}\omega \tau},  
\end{equation}
and ${\tilde {D}}_{\bf \mu}$ denotes the gauge-invariant gradient on the lattice accompanied by the gauge field $\delta {\bf A}(\tau_1) - \delta {\bf A}(\tau_2)$. Performing the cumulant expansion in powers of $q^{(ab)}$ and $\psi^{(a)}$ in the logarithmic term and replacing\cite{Read} $q^{ab}(\tau_1,\tau_2)$ by $Q^{ab}(\tau_1,\tau_2) - C \delta_{a,b} \delta(\tau_1-\tau_2)$ with a constant $C$, we finally obtain the following effective Landau action 
\begin{eqnarray}\label{eq:12}
t {\cal S}_{\rm eff}(\psi, Q; \delta {\bf A}) &=&  \int \frac{d^d{\bf r}}{a^d} \biggl[ \int \frac{d\tau}{\kappa} \sum_a \biggl( \frac{\partial^2}{\partial \tau_1 \, \partial \tau_2} + r \biggr)  Q^{(aa)}({\bf r}; \tau_1, \tau_2) \biggr|_{\tau_1=\tau_2} \nonumber \\ 
&-& \frac{\kappa}{3} \int d\tau_1 d\tau_2 d\tau_3 \sum_{a,b,c} Q^{(ab)}({\bf r}; \tau_1,\tau_2) Q^{(bc)}({\bf r}; \tau_2,\tau_3) Q^{(ca)}({\bf r}; \tau_3, \tau_1) \nonumber \\
&+& \frac{t a^2}{4 d \alpha^2} \sum_{a,b} \int d\tau_1 \int d\tau_2 |(-{\rm i}\nabla_{\bf r} - e^*(\delta {\bf A}(\tau_1) - \delta {\bf A}(\tau_2))) Q^{(ab)}({\bf r}; \tau_1, \tau_2)|^2 \nonumber \\
&+& \frac{u}{2} \int d\tau \sum_a (Q^{(aa)}({\bf r}; \tau,\tau))^2  \biggr] + t {\tilde {\cal S}}_{\rm eff}, 
\end{eqnarray}
where 
\begin{eqnarray}\label{eq:13}
t {\tilde {\cal S}}_{\rm eff} &=& a^{-d} \int d^d{\bf r} \biggl[ \sum_a \biggl(  \int d\tau \biggl[ r_{\psi, 0} |\psi^{(a)}(\tau)|^2 \nonumber \\
&+& c_\psi \biggl|\frac{\partial \psi^{(a)}}{\partial \tau} \biggr|^2 
+ t \, {\tilde a}^2 |(-{\rm i}\nabla_{\bf r} - e^* {\bf A}_{\rm ex} - e^* \delta {\bf A}(\tau) ) \psi^{(a)}(\tau)|^2 \\ \nonumber 
&+& \frac{t}{2 \alpha} \biggl(\frac{u_R}{\alpha} \biggr) \biggl(\frac{4 J_0}{\alpha} \biggr)^2 |\psi^{(a)}({\bf r}, \tau)|^4 \biggr ] \biggr) \\ \nonumber
&-& \frac{t}{8} \sum_{a,b} \int d\tau_1 \int d\tau_2 \, \frac{(2J/\alpha)^4}{\sqrt{(1+(2J/\alpha)^2)(1 - 3(2J/\alpha)^2)}} \, |\psi^{(a)}({\bf r}, \tau_1)|^2 \, |\psi^{(b)}({\bf r}, \tau_2)|^2 
 \biggr].
\end{eqnarray}
Further, the ${\bf r}$-sum was transformed into the space integral, and instead, a short length cut-off $a$ was introduced. Note that the $\delta {\bf A}$-dependent term arises from a term bilinear in $q^{(ab)}(\tau_1,\tau_2)$, and a $Q^{(aa)}$-linear term carries no $\delta {\bf A}$. 

We note that, although the dissipative term in eq.(\ref{eq:1}) directly appears only in the term quadratic in $\psi$ of the effective action, it also affects the dynamics of the glass field $Q^{(ab)}$ through the coupling ($w_\psi$-) term between the two fields. 

The above expression of the action is of the same form as those in other works \cite{DP,PD}. As explicitly examined in Ref.\cite{RIsub}, the coefficients $r$, $u$, $t$, $\kappa$, $c_\psi$, $d_\psi$, and $w_\psi$ are positive, while 
\begin{eqnarray}\label{eq:15}
\frac{r_{\psi,0}}{t} &=& \frac{1}{4d} - \frac{J_0}{2 \alpha} + \frac{J_0 T}{2 \alpha^2}- \frac{J_0}{4 \alpha} \biggl( \biggl(\frac{\alpha}{2J} \biggr)^2 - 1 \biggr), \\ \nonumber 
\cr {\tilde a}^2 &=& a^2 \frac{J_0}{4 d \alpha}. 
\end{eqnarray} 
Below, the Fourier transform of the glass field $Q^{(ab)}({\bf r})$ is defined, by following Ref.\cite{Read}, as 
\begin{eqnarray}\label{eq:16}
Q^{(ab)}({\bf r}; \tau_1,\tau_2) 
&=& q^{(ab)} 
 + \frac{1}{\beta} \sum_{\omega \neq 0} {\overline D}_\omega e^{-{\rm i}\omega(\tau_1-\tau_2)} \, \delta_{a,b} \\ \nonumber 
&+& \beta^{-2} \sum_{\omega_1, \omega_2} \delta Q^{(ab)}_{\omega_1, \omega_2}({\bf r}) e^{-{\rm i}\omega_1 \tau_1 - {\rm i}\omega_2 \tau_2}, 
\end{eqnarray}
where 
\begin{equation}\label{eq:rs}
q^{(ab)} = q (1 - \delta_{a,b}) + {\overline q} \, \delta_{a,b}
\end{equation}
in the replica-symmetric approximation adopted in the previous \cite{Read,DP,WP} papers. 

Now, we examine the conductivity in PG at low enough $T$ by, as in other works \cite{Read,DP,PD,WP}, treating the PG order parameter field in the mean field (MF) approximation where $\delta Q^{(ab)}({\bf r})=0$. We note that, in eq.(\ref{eq:13}), the $\psi$-field (i.e., SC fluctuation) couples to the glass field $Q^{(ab)}$ in a bilinear form $(\psi^{(a)})^* \psi^{(b)}$ of which the diagonal ($a=b$) component is nothing but the SC fluctuation contribution to the entropy density irrespective of the presence or absence of the applied field. Thus, a result on the conductivity in zero fields ($H=0$) is trivially extended to the $H > 0$ case. In other words, the metallic conductivity \cite{DP,PD} in PG, if correct, would also affect the $H$-$T$ phase diagram at low enough $T$. Bearing this in mind, we will discuss, for the moment, results in the simpler $H=0$ case with ${\bf A}_{\rm ex}=0$ and concentrate on results valid for both the $H=0$ and $H \neq 0$ cases. Our analysis is different from that in other works \cite{DP,PD} in that we take account of a dissipative dynamics correctly and of the presence of the SC fluctuation {\it consistently} in determining the MF solution of $Q^{(ab)}$. As shown below, the coupling between the SC fluctuation and the MF $Q^{(ab)}$ makes the conductivity in the PG ordered state not finite but divergent and hence, makes the PG a superconducting phase. 

Deep in the PG, it is sufficient to keep the $\psi$-fluctuation in the Gaussian approximation. In $H \neq 0$ case, the corresponding approximation may be valid {\it above} $H_{c2}(0)$ and at low enough temperatures \cite{JL}. Below, the replica symmetry will be assumed even for the static ($\omega=0$) components on the basis of the argument in the literature \cite{Read} that taking the replica symmetry is justified in low $T$ limit. A breaking of replica symmetry cannot qualitatively affect our main conclusions given below, because it would be accompanied by an {\it independent} parameter such as a coefficient of a quartic term 
on $Q^{ab}$ in the effective action. 
Then, using eq.(\ref{eq:16}) and neglecting $\delta Q^{(ab)}$, the variational equation $0= {\rm lim}_{n \to +0} n^{-1} \partial {\overline {Z^n}}/\partial Q^{(ab)}(\tau_1,\tau_2)$ takes the form of the following three equations: 
\begin{eqnarray}\label{eq:17}
\kappa^{-1}(\omega^2+r) &-& \kappa {\overline D}_\omega^2 + u({\overline q}+\beta^{-1}\sum_{\omega \neq 0} {\overline D}_\omega) \\ \nonumber 
&-& w_\psi \int_{\bf k} G_{\rm dia}^{(d)}({\bf k}, \omega)=0 ,
\end{eqnarray}
for nonzero $\omega$, 
\begin{eqnarray}\label{eq:18}
\kappa^{-1} r &-& \kappa \beta^2 ({\overline q}^2 - q^2) + u({\overline q}+\beta^{-1} \sum_{\omega \neq 0} {\overline D}_\omega) \\ \nonumber 
&-& w_\psi \int_{\bf k} G_{\rm dia}^{(d)}({\bf k}; 0)=0, 
\end{eqnarray}
\begin{equation}\label{eq:19}
2 \kappa \beta^2 q ({\overline q}-q) + w_\psi \int_{\bf k} G_{\rm od}^{(d)}({\bf k}; 0) 
= 0, 
\end{equation}
where $\int_{\bf k}$ denotes $\int d^d(ka)/(2 \pi)^d$. These are obtained as the variational equations on ${\overline D}_\omega$, ${\overline q}$, and $q$, respectively. Here, we have expressed the SC fluctuation propagator $G_{ab}^{(d)}({\bf k}; \omega)= \beta^{-1}<(\psi_i^{(a)}({\bf k}; \omega))^* \psi_i^{(b)}({\bf k}; \omega)>$ in $d$-dimension in the form $G_{ab}^{(d)}({\bf k}; \omega) = \delta_{a,b}(1-\delta_{\omega,0}) G_{\rm dia}^{(d)}({\bf k}; \omega) + \delta_{\omega,0} [ \, \delta_{a,b} G_{\rm dia}^{(d)}({\bf k}; 0) + (1-\delta_{a,b})G_{\rm od}^{(d)}({\bf k}; 0) \, ]$, 
where
\begin{eqnarray}\label{eq:20}
G_{\rm dia}^{(d)}({\bf k}; \omega) &=& \frac{t}{r_\psi+d_\psi |\omega| + c_\psi \omega^2 + t {\tilde a}^2 k^2 - w_\psi {\overline D}_\omega}, \\ \nonumber
\cr G_{\rm od}^{(d)}({\bf k}; 0)&=&\frac{w_\psi \beta q}{t} (g_{\bf k}(\Delta q))^2, \\ \nonumber 
\cr G_{\rm dia}^{(d)}({\bf k}; 0)&=& g_{\bf k}(\Delta q) + G_{\rm od}^{(d)}({\bf k}; 0), \\ \nonumber
\cr g_{\bf k}(\Delta q) &=& \frac{t}{r_\psi + t {\tilde a}^2 k^2 + w_\psi \Delta q}, 
\end{eqnarray}
and 
\begin{equation}\label{eq:21}
\Delta q= \beta (q-{\overline q}). 
\end{equation} 
Noting that, when $q > 0$, eq.(\ref{eq:19}) becomes 
\begin{equation}\label{eq:22}
\Delta q=\biggl(\frac{w_\psi}{t} \biggr)^2 \frac{t}{2 \kappa} \int_{\bf k} (g_{\bf k}(\Delta q))^2, 
\end{equation}
we easily find that the only physically meaningful solution of the PG order parameter is given together with eq.(\ref{eq:22}) by 
\begin{eqnarray}\label{eq:23}
{\overline D}_\omega &=& - \Delta q - \kappa^{-1}  \, p_\psi \, 
|\omega|^{1/2}, \\ \nonumber 
{\overline q} &=& - \beta^{-1} \sum_{\omega \neq 0} {\overline D}_\omega + u^{-1} (\kappa (\Delta q)^2 + w_\psi \int_{\bf k} g_{\bf k}(\Delta q) \\ \nonumber 
&-& \kappa^{-1} r), 
\end{eqnarray}
where 
\begin{equation}
p_\psi = \biggl(\frac{\kappa d_\psi w_\psi t^{-1} \int_{\bf k} (g_{\bf k})^2}{1 + t \kappa^{-1} \int_{\bf k} (g_{\bf k})^3 (w_\psi  \, t^{-1} )^3} 
\biggr)^{1/2}.
\end{equation}
The above form of ${\overline D}_\omega$ is valid up to O($|\omega|^{1/2}$). When the dissipative term in $G_{\rm dia}^{(d)}({\bf k}, \omega)$ is absent, the $-|\omega|^{1/2}$ term in eq.(24) is replaced by a $-|\omega|$ term. In $w_\psi \to 0$ limit where $\psi$ and $Q$ fields are decoupled, the above MF solution reduces to the one \cite{Read,PD} $- \kappa^{-1} |\omega|$ with $q > 0$ and $\psi=0$ if higher order terms, omitted in eq.(\ref{eq:23}), are kept. Note that the $|\omega|^{1/2}$-term, arising from the dissipative dynamics in eq.(\ref{eq:2}), of ${\overline D}_\omega$ was brought by the $\psi$-fluctuation. Consistently, this term appears in $G^{(d)}_{\rm dis}({\bf k}; \omega)$ and makes the dynamics of $\psi$-fluctuation in the PG state sub-Ohmic. 

The analysis in zero field case given above is applied to granular superconductors in {\it nonzero} fields in the following way: As in eq.(\ref{eq:22}), effects of SC fluctuation on the glass order are included merely in a form of an integral over the momentum ${\bf k}$ of $(g_{\bf k})^m$ or $G_{\rm dia}^{(d)}({\bf k}; \omega)$. In nonzero fields ($H > 0$), the $\psi$-field is decomposed into the Landau levels, and one has only to replace the ${\bf k}_\perp$-integral and $k_\perp^2$ by a summation over the Landau level index $l$ and $|e^*| H (2l+1)$, respectively, where $e^*$ is the charge of Cooper pairs. Thus, it is easily understood that eq.(\ref{eq:23}) and the divergent conductivity given below are also valid in $H \neq 0$ case if such replacements are performed. 

Further, we note that the dissipative term, eq.(4), may be alternatively incorporated as follows: To simplify our description, any spatial variation of the glass order parameter $q$ will not be considered so that the gradient ${\tilde D}_{\bf \mu}$ may be negligible. Further, the dissipative term will be treated here in the form not of ${\cal S}_f$ but of ${\tilde {\cal S}}_g \equiv {\cal S}_{\rm dis} + {\cal S}_g$. Explicitly, 
\begin{eqnarray}
{\tilde {\cal S}}_g \! &=& \! \sum_{\bf r} \biggl( \frac{J^{-2}}{2} \int d\tau \int d\tau' \sum_{a,b} q^{(ba)}(\tau', \tau) \, q^{(ab)}(\tau, \tau') \nonumber \\
&-& \, {\rm ln} \langle T_\tau \exp\biggl[ - \beta^{-1} \sum_\omega \sum_a |\omega| |\Phi^{(a)}_{\bf r}(\omega)|^2 
+ \frac{1}{2} \sum_{a,b} \int d\tau \int d\tau' {\overline \Phi}^{(ab)}_{\bf r}(\tau, \tau') q^{(ba)}(\tau', \tau) + {\rm c.c.} \biggr] \rangle \biggr),
\end{eqnarray}
where ${\overline \Phi}^{(ab)}_{\bf r}(\tau, \tau')=\Phi^{(a)}_{\bf r}(\tau) (\Phi^{(b)}_{\bf r}(\tau'))^*$. Here, let us focus on the replica diagonal terms. Then, expressing $q^{(aa)}(\tau, \tau')$ as $\beta^{-1} \sum_\omega d_\omega \exp(-{\rm i}\omega(\tau-\tau'))$ and replacing $d_\omega - |\omega|$ 
by $d_\omega$, the above expression is rewritten as 
\begin{eqnarray}
{\tilde {\cal S}}_g \! &=& \! \sum_{\bf r} \biggl( J^{-2} \sum_\omega |\omega| d_\omega + \frac{J^{-2}}{2} \int d\tau \int d\tau' \sum_{a,b} q^{(ba)}(\tau', \tau) \, q^{(ab)}(\tau, \tau') \nonumber \\ 
&-& \, {\rm ln} \langle T_\tau \exp\biggl[\frac{1}{2} \sum_{a,b} \int d\tau \int d\tau' {\overline \Phi}^{(ab)}(\tau, \tau') q^{(ba)}(\tau', \tau) + {\rm c.c.} \biggr] \rangle \biggr). 
\end{eqnarray}
Except the first term, the above expression is the same form as that in the nondissipative case. Contrary to the claim \cite{WP}, the first term is {\it not} lost after the replacement $q^{(ab)}(\tau, \tau') \to Q^{(ab)}(\tau, \tau') - C \delta_{a,b} \delta(\tau - \tau')$. That is, the $|\omega|$ term is added to the first ($\omega^2$-) term in eq.(18). Then, it will be obvious that the variational solution of $Q^{(aa)}$ takes the form $\propto - |\omega|^{1/2}$ as its leading $\omega$ dependence at low $|\omega|$, just as in eq.(24) (see the sentences following eq.(25)). In Ref.\cite{WP}, the first term was erroneously omitted, and consequently, the argument of the metallic glass phase was not changed. 

Now, let us examine the conductivity $\sigma$ following from the solution, eq.(25), in terms of Kubo formula. When the glass fluctuation $\delta Q^{(ab)}$ is neglected, the conductivity following from the present model consists of two parts. One is the direct contribution $\sigma_\psi$ from the SC fluctuation and, as usual, has the form 
\begin{eqnarray}\label{eq:AL}
\sigma_\psi(\Omega) &=& \frac{4 (e^*)^2}{|\omega|} \, \frac{{\tilde a}^2}{a^2} \int_{\bf k} \frac{k^2}{d} \beta^{-1} \sum_{\omega_1} \frac{1}{n} \sum_{a,b} G^{(d)}_{ab}({\bf k}; \omega_1) \\ \nonumber 
&\times& [ G^{(d)}_{ab}({\bf k}; \omega_1) - G^{(d)}_{ab}({\bf k}; \omega_1 + \omega) ] \biggr|_{{\rm i}\omega \to \Omega+{\rm i}0}, 
\end{eqnarray}
where the $n \to 0$ limit is taken at the end. In the disordered (i.e., normal) phase in which $q=0$, this expression was examined previously and shown to vanish in $T \to 0$ limit \cite{RIr} although the bare dynamics is dissipative. In the PG phase, by substituting eqs.(\ref{eq:20}) with (\ref{eq:23}) into eq.(\ref{eq:AL}), $\sigma_\psi$ becomes 
\begin{equation}\label{eq:101}
{\rm Re} \sigma_\psi(\Omega \to 0) = \frac{2 \sqrt{2} q (e^*)^2 \, p_\psi}{\kappa \, |\Omega|^{1/2}} \biggl(\frac{w_\psi}{t} \biggr)^2 \int_{\bf k} k^2 (g_{\bf k})^3 
\end{equation}
for $d=2$, implying a divergent contribution to the dc conductivity arising from the sub-Ohmic dynamics of the SC fluctuation in the PG phase. Note that the $T \to 0$ limit was not taken in obtaining eq.(\ref{eq:101}).  

To corroborate that the total conductivity is also divergent, let us examine another contribution 
\begin{eqnarray}\label{eq:24}
\sigma_{\rm PG}({\rm i}\omega) &=& \frac{1}{|\omega| \, d} \biggl(\frac{e^*}{\alpha} \biggr)^2 \lim_{n \to 0} \frac{1}{n} \sum_{a,b} \int_0^\beta d \tau_1 e^{{\rm i}\omega(\tau_1-\tau_3)} \\ \nonumber 
&\times& \biggl[ \int_0^\beta d\tau_2 \delta(\tau_1-\tau_3) |Q^{(ab)}(\tau_1, \tau_2)|^2 - |Q^{(ab)}(\tau_1-\tau_3)|^2 \biggr].
\end{eqnarray}
to the conductivity which arises from the terms (see eq.(\ref{eq:12})) quadratic in $Q^{(ab)}$ and dependent on $\delta {\bf A}$. 
By substituting eq.(\ref{eq:16}) into eq.(\ref{eq:24}), it becomes 
\begin{eqnarray}\label{eq:25}
\sigma_{\rm PG}({\rm i} \omega) &=& \frac{1}{|\omega| \, d} \biggl(\frac{e^*}{\alpha} \biggr)^2 \, (1 - \delta_{\omega, 0}) \, \biggl[ \frac{\beta}{n} \sum_{a,b} q^{(ab)} q^{(ba)} + \beta^{-1} \sum_{\omega_1} {\overline D}_{\omega_1} ({\overline D}_{\omega_1} - {\overline D}_{\omega_1+\omega}) \\ \nonumber 
\cr &+& \beta^{-1} (2 {\overline D}_\omega - {\overline D}_0) {\overline D}_0 - 2 q^{(aa)} \, {\overline D}_{\omega} \biggr] \\ \nonumber 
&=& \frac{1}{|\omega| \, d} \biggl(\frac{e^*}{\alpha} \biggr)^2 \, \biggl[ (1 - \delta_{\omega, 0}) \, \frac{\beta}{n} \sum_{a,b} {\tilde q}^{(ab)} {\tilde q}^{(ba)} + \frac{\beta^{-1}}{2} \sum_{\omega_1} ({\overline D}_{\omega_1} - {\overline D}_{\omega_1+\omega})^2 
+ 2q ({\overline D}(0) - {\overline D}(\omega)) \biggr], 
\end{eqnarray}
where the replica-symmetric expression (\ref{eq:rs}) of $q^{(ab)}$ was used in obtaining the expression following the second equality. Consequently, we have 
${\tilde q}^{(ab)} = q$, and the sum ${\tilde q}^{(ab)} {\tilde q}^{(ba)}$ term is absent in $n \to 0$ limit. Then, let us turn to the terms including ${\overline D}_\omega$. The real part of the term under the $\omega_1$-summation is positive and nondivergent after the substitution ${\rm i}\omega \to \Omega$. It can be easily seen in terms of the identity 
\begin{equation}
|\omega|^{1/2} = \pi^{-1} \int dx \frac{|\omega|}{x^2 + |\omega|}.
\end{equation}
In contrast, according to eq.(\ref{eq:23}), the term proportional to $q$ is positive and, as well as eq.(\ref{eq:101}), becomes divergent like $|\Omega|^{-1/2}$ in dc limit. The above results altogether imply that the PG phase is {\it superconducting}. We stress that this result has been obtained by including the coupling between the SC ($\psi$-) fluctuation and the glass order parameter $Q^{(ab)}$ in obtaining the MF $Q^{(ab)}$-solution.

\section{III. Extension to glass phase due to line-like correlated disorder}

Here, our analysis of the conductivity will be extended to the case of the so-called BG state which occurs due to line defects persistent along the applied field $\parallel {\hat z}$. Then, we assume the starting model eq.(2) to, in turn, have the correlation 
${\overline {J_{\bf \mu}({\bf r}) J^*_{{\bf \mu}'}({\bf r}')}} = (4 J^2/d) \delta_{{\bf \mu}, {\bf \mu}'} \delta_{{\bf x}, {\bf x}'}$ 
of the random Josephson coupling, where ${\bf r}=({\bf x}$, $z)$, ${\bf r}'=({\bf x}'$, $z')$, and ${\bf x}$ (${\bf x}'$) is the in-plane component of the coordinate ${\bf r}$ (${\bf r}'$). To point out our main point that, with no dissipative dynamics, the physically acceptable response properties in a glass phase do not follow, we only have to focus here on the simple case with no coupling to the $\psi$-fluctuation and with no dissipative phase dynamics. If incorporating the gauge field $\delta {\bf A} \perp {\hat z}$ necessary in examining the conductivity perpendicular to the field, the replicated action related to the glass field $q^{(ab)}$ or $Q^{(ab)}$ takes the form
\begin{eqnarray}
{\cal S}_g \! &=& \! - \frac{J^2}{8d} \int d\tau \int d\tau' \sum_{a,b} \sum_{\bf x} \sum_{{\bf \mu}} \sum_{z,z'} \biggl[ \, \, {\overline \Phi}_{\bf x}^{(ba)}(z', z; \tau', \tau) \, {\overline \Phi}_{\bf x}^{(ab)}(z+\mu a, z'+ \mu a; \tau, \tau') \nonumber \\ 
&\times& \exp({\rm i}(\delta A_\mu(\tau) - \delta A_\mu(\tau'))) + {\rm c.c.} \, \, \biggr] \nonumber \\
&=& \! \! \frac{- J^2}{2} \int \! d\tau \! \int d\tau' \! \sum_{a,b} \sum_{{\bf x}, z,z'} \biggl[ {\overline \Phi}^{(ba)}_{\bf x}(z', z; \tau', \tau) \biggl( 1 + \frac{{\tilde D}_\mu \cdot {\tilde D}_\mu^*}{2d} \biggr) {\overline \Phi}^{(ab)}_{\bf x}(z, z'; \tau, \tau') + {\rm c.c.} \biggr],
\end{eqnarray}
where ${\overline \Phi}_{\bf x}^{(ba)}(z', z; \tau', \tau) = \Phi_{\bf x}^{(b)}(z', \tau') \, (\Phi_{\bf x}^{(a)}(z, \tau))^*$, and $\Phi_{\bf x}^{(a)}(z, \tau) = \exp({\rm i} \theta^{(a)}({\bf x}, z; \tau))$. Then, by assuming the cases with the glass field independent of ${\bf x}$, the effective action 
corresponding to eq.(11) and with no $\psi$-field becomes 
\begin{eqnarray}
{\cal S}_{\rm eff} &=& \frac{J^{-2}}{2} \sum_{\bf x} \int d\tau \int d\tau' \sum_{a,b} \sum_{z,z'} q^{(ba)}_{\bf x}(z',z; \tau', \tau) \, q^{(ab)}_{\bf x}(z,z'; \tau, \tau') \nonumber \\ 
&-& \sum_{\bf x} \, {\rm ln} \langle T_\tau \exp\biggl[\frac{1}{2} \sum_{a,b} \sum_{z,z'} \int d\tau \int d\tau' {\overline \Phi}_{\bf x}^{(ab)}(z, z'; \tau, \tau') \biggl( 1 - \frac{a^2}{4d} \frac{\partial^2}{\partial z^2} \biggr) q^{(ba)}_{\bf x}(z', z; \tau', \tau) + {\rm c.c.} \biggr] \rangle \nonumber \\
&=& \sum_{\bf x} \biggl[ - \sum_a \sum_{z,z'} \int d\tau \int d\tau' \delta_{z,z'} G(\tau-\tau') \biggl( 1 - \frac{a^2}{4d} \frac{\partial^2}{\partial z^2} \biggr) q^{(aa)}(z', z; \tau', \tau) \nonumber \\ 
&+& \frac{1}{2} \sum_{a,b} \prod_{j=1}^{4} \sum_{z_j} \int d\tau_j \biggl(J^{-2} \delta(\tau_1-\tau_4) \delta(\tau_2-\tau_3) - G(\tau_1-\tau_4) G(\tau_2-\tau_3) \biggr) \delta_{z_1, z_4} \delta_{z_2, z_3} \, q^{(ab)}_{12} q^{(ba)}_{34} \nonumber \\
&-& \! \frac{1}{3} \sum_{a,b,c} \prod_{j=1}^6 \sum_{z_j} \int d\tau_j G(\tau_1-\tau_6) G(\tau_2-\tau_3) G(\tau_4-\tau_5) \delta_{z_1,z_6} \delta_{z_2,z_3} \delta_{z_4,z_5} q^{(ab)}_{12} q^{(bc)}_{34} q^{(ca)}_{56} 
+ \cdot \cdot \cdot \biggr], 
\end{eqnarray}
where $q^{(ab)}_{mn} = q^{(ab)}(z_m, z_n; \tau_m, \tau_n)$. After the transformation into the $Q^{(ab)}$-field, this action becomes 
\begin{eqnarray}
t {\cal S}_{\rm eff} &=& \sum_{\bf x} \biggl[ \sum_{z_1, z_2} \sum_a \kappa^{-1} \int d\tau_1 \int d\tau_2 \delta(\tau_1 - \tau_2) \delta_{z_1, z_2} \biggl( \frac{\partial^2}{\partial \tau_1 \partial \tau_2} + r + \frac{\alpha^2 a^2}{4d} \frac{\partial^2}{\partial z_1 \partial z_2} \biggr) Q^{(aa)}_{12}({\bf x}) \nonumber \\
&-& \sum_{a,b,c} \frac{\kappa}{3} \prod_{j=1}^3 \int d\tau_j \sum_{z_j} Q^{(ab)}_{12}({\bf x}) Q^{(bc)}_{23}({\bf x}) Q^{(ca)}_{31}({\bf x}) + u \int d\tau_1 \sum_{z_1} (Q^{(aa)}_{11})^2 \nonumber \\
&+& \sum_{a,b} \sum_{z_1, z_2} \int d\tau_1 \int d\tau_2 \biggl[ \frac{t a^2 (e^*)^2}{4 d \alpha^2}  \sum_{\mu \neq \pm z}(\delta {A}_\mu(\tau_1) - \delta {A}_\mu(\tau_2))^2 Q^{(ab)}_{12} Q^{(ba)}_{21} \nonumber \\
&-& \frac{J_0}{\alpha^2} (\psi^{(a)}_1)^* Q^{(ab)}_{12} \psi^{(b)}_2 \biggr] \biggr]. 
\end{eqnarray}
This form of the effective action implies that the nonlocality in the $z$-direction $\parallel {\bf H}$ appears in parallel with that in the imaginary time. Hence, if the glass fluctuation is negligible, the glass field $Q^{(ab)}_{12}$ should take the form 
\begin{equation}
Q^{(ab)}_{12} \equiv Q^{(ab)}(z_1, z_2; \tau_1, \tau_2) = q^{(ab)} + \beta^{-1} L_z^{-1} \sum_{\omega, k} D_{\omega}(k_z) \delta_{a,b} \exp({\rm i} k_z (z_1-z_2)- {\rm i} \omega (\tau_1 - \tau_2))
\end{equation}
as an extension of that in the point disorder case. Then, it is straightforward to verify that, in the replica symmetric approximation, 
\begin{eqnarray}
q_{ab} &=& q, \\ \nonumber
D_{\omega}(k_z) &=& -\kappa^{-1} \sqrt{ \omega^2 + \alpha^2 a^2 k_z^2/4}. 
\end{eqnarray}

This form of the $Q^{(ab)}$-solution indicates that, as far as focusing on the Kubo formula of the conductivity, contributions to the conductivity from the $k_z=0$ components of $Q^{(aa)}$ should be the same, in the BG phase, as those in the 2D VG point phase at $T=0$. To see this, let us first focus on the counterpart of eq.(32). In the case of line disorder, it is easily seen that, using eq.(37), the counterpart of eq.(32) becomes 
\begin{eqnarray}\label{eq:38}
\sigma_{\rm PG}^{({\rm BG})}({\rm i}\omega) &=& \frac{1}{|\omega| \, d} \biggl(\frac{e^*}{\alpha} \biggr)^2 \lim_{n \to 0} \frac{1}{n} \sum_{z_1,z_2} \sum_{a,b} \int_0^\beta d \tau_1 e^{{\rm i}\omega(\tau_1-\tau_3)} \\ \nonumber 
&\times& \biggl[ \int_0^\beta d\tau_2 \, \delta(\tau_1-\tau_3) |Q^{(ab)}_{12}|^2 - |Q^{(ab)}_{13}|^2 \biggr],
\end{eqnarray}
and that the terms proportional to $q$ (see the final term of eq.(33)) in eq.(40), which are the only contributions there leading to a nonzero conductivity in $T \to 0$ limit, consist only of the $k_z=0$ contribution. Thus, just as in the nondissipative point disorder case, a nonzero positive contribution to the conductivity follows from eq.(\ref{eq:38}). Further, the contribution, eq.(31), related to the SC fluctuation also gives a positive contribution to the conductivity stemming from $D_{\omega}(k_z=0)$ and is expressed by 
\begin{equation}
\sigma_\psi^{({\rm BG})}(i \omega) = \frac{d \sqrt{2} q (e^*)^2}{|\omega|} \, \biggl(\frac{w_\psi}{t} \biggr)^2 \int_{{\bf k}_\perp} k_\perp^2 (-D_{\omega}(0)) (g_{{\bf k}_\perp, k_z=0})^3. 
\end{equation}
Again, this SC fluctuation contribution is also finite in the nondissipative case. Therefore, the results on the conductivity in BG phase are qualitatively the same as in VG phase. By repeating the analysis mentioned above in the case with dissipative dynamics, it is easy to verify that the contributions to the conductivity corresponding to eqs.(40) and (41) are divergent in the dc limit $|\Omega| \to 0$ in the case with dissipative dynamics. 

\section{IV. Summary}

First, we have shown by applying a theory \cite{Read} of quantum spin-glass to the quantum granular superconductor, modelled by eqs.(\ref{eq:1}) and (\ref{eq:2}), that, even if assuming the replica symmetry, a PG ordered state peculiar to granular superconductors is characterized by zero resistance when the Ohmic dissipative phase dynamics is incorporated. This is consistent not with the argument \cite{DP} of a metallic response in this phase but with the conventional picture \cite{GaL,JL} of a {\it superconducting} glass existing in $H > H_{c2}(0)$ (see Fig.1). Elsewhere, we have shown \cite{RIsub} that, even without the PG order, the fluctuation of the PG order parameter plays the role of {\it pinning disorder} inducing a VG instability at $T > 0$, at which the resistivity continuously vanishes, in 3D and nonzero field case. The present result is consistent with results of the approach \cite{RIsub} from higher temperatures. 

We have also examined the case with the glass (BG) phase due to line-like disorder in addition to the familiar VG case due to point disorder in order to corroborate that the argument \cite{WP} that the VG is a metal is false even from the viewpoint favoring consistencies with experimental facts. It was found that, even in the case with line defects, the conductivity in the resulting BG phase also becomes nondivergent when the phase dynamics is nondissipative, while it is divergent in dc limit when the Ohmic dissipation is introduced. Thus, the argument in Ref.\cite{WP} based on the nondissipative model that the VG is a metal is equivalent to concluding that even the BG is also a metal, contrary to the well accepted experimental fact that the BG has zero dc resistance, and hence, has no reliable basis both theoretically and experimentally. The result that, in both cases of point-like and line-like disorders, the glass phase becomes superconducting in the presence of dissipative dynamics indicates that there is no theoretical basis of resulting in an intermediate bose-metal phase in disordered granular systems at low $T$ limit. When trying to understand superconducting resistive behaviors upon cooling, inclusion of a dissipative dynamics is essential even in the quantum regime \cite{RIr}. 

This work is finantially supported by a Grant-in-Aid for Scientific Research on Priority Areas from MEXT, Japan. 

\vspace{5mm}



\begin{thebibliography}{99}

\bibitem{DP} D. Dalidovich and P. Phillips, Phys. Rev. Lett. {\bf 89} (2002) 027001. 
\bibitem{PD} P. Phillips and D. Dalidovich, Phys. Rev. B {\bf 69} (2003) 104427 . 
\bibitem{Aharon} N. Mason and A. Kapitulnik, Phys. Rev. B {\bf 65} (2002) 220505(R). 
\bibitem{Vicente} Y. Qin, C. L. Vicente, and J. Yoon, Phys. Rev. B {\bf 73} (2006) 100505(R). 
\bibitem{FFH} D. S. Fisher, M. P. A. Fisher, and D. A. Huse, Phys. Rev. B {\bf 43} (1991) 130. 
\bibitem{Feigel} M.V. Feigel'man and L.B. Ioffe, Phys. Rev. Lett. {\bf 74} (1995) 3447 and L.B. Ioffe, Phys. Rev. B {\bf 38} (1988) 5181. 
\bibitem{WP} I. Wu and P. Phillips, Phys. Rev. B {\bf 73} (2006) 214507. 
\bibitem{RIsub} R. Ikeda, Phys. Rev. B {\bf 74} (2006) 054510. 
\bibitem{Kleinert} H. Kleinert, {\it Gauge Fields in Condensed Matter} (Word Scientific, 1989). 
\bibitem{Read} N. Read, S. Sachdev, and J. Ye, Phys. Rev. B {\bf 52} (1995) 
384. 
\bibitem{GaL} V. M. Galitski and A. I. Larkin, Phys. Rev. Lett. {\bf 87} (2001) 087001. 
\bibitem{JL} S. John and T. C. Lubensky, Phys. Rev. B {\bf 34} (1986) 4815.
\bibitem{RIr} R. Ikeda, Int. J. Mod. Phys. B {\bf 10} (1996) 601. 
\end{thebibliography}
\end{document}